%% file: minimalistic_widening.tex
\documentclass[a4paper]{article}
\usepackage{ifpdf}
\usepackage{amsfonts,amssymb,stmaryrd}
\usepackage[sort]{natbib}
\usepackage{amsthm}
\usepackage{graphicx,color}
\usepackage{hyperref}

\author{David Monniaux\\\small CNRS / VERIMAG%
\thanks{VERIMAG is a joint laboratory of CNRS, Universit\'e Joseph Fourier and Grenoble-INP.}}
\title{A minimalistic look at widening operators}

\newcommand{\parts}[1]{\mathcal{P}(#1)}
\newcommand{\abstr}[1]{{#1}^\sharp}
\newcommand{\bbZ}{\mathbb{Z}}
\newcommand{\bbQ}{\mathbb{Q}}
\newcommand{\bbN}{\mathbb{N}}
\newcommand{\widening}{\triangledown}

\theoremstyle{definition}
\newtheorem{defn}{Definition}

\ifpdf

\else

\fi

\begin{document}
\maketitle

\begin{abstract}
We consider the problem of formalizing in higher-order logic the familiar notion of widening from abstract interpretation. It turns out that many axioms of widening (e.g. widening sequences are ascending) are not useful for proving correctness. After keeping only useful axioms, we give an equivalent characterization of widening as a lazily constructed well-founded tree. In type systems supporting dependent products and sums, this tree can be made to reflect the condition of correct termination of the widening sequence.
\end{abstract}

\section{The usual framework}
\label{part:absint}
We shall first recall the usual definitions of abstract interpretation and widening operators.

\subsection{Abstraction and concretization maps}
Abstract interpretation is a framework for formalizing approximation relationships arising in program semantics and static analysis \citep{Cousot_state_doctorate,CousotCousot_JLC92}. \emph{Soundness} of the abstraction is expressed by the fact that the approximation takes place in a controlled direction. In order to prove that a given set of undesirable states is unreachable, we can compute a superset of the set of reachable states (an \emph{over-approximation} thereof), in the hope that this set does not intersect the set of undesirable states. If order to prove that we eventually reach a given set of states, we can compute a subset of the set of states that eventually reach them (an \emph{under-approximation} thereof), in the hope that this set includes the initial states.

Most introductory materials on abstract interpretation describe abstraction as a \emph{Galois connection} between a concrete space $S$ (typically, the powerset $\parts{\Sigma}$ of the set of states $\Sigma$ of the program, or the powerset of the set of finite execution traces $\Sigma^*$ of the program) and an abstract space $\abstr{S}$. For instance, if the program state consists in a program counter location, taken within a finite set $P$ of program locations, and three integer variables, $\Sigma = P \times \bbZ^3$, $S = \parts{P \times \bbZ^3}$, the abstract state can be, for instance, a member of $\abstr{S} = P \rightarrow (\{ \bot \} \cup I^3)$, where $P$ is the set of program locations, $a \rightarrow b$ denotes the set of functions from $a$ to $b$, $I$ is the set of well-formed pairs $(a,b)$ defining intervals ($a \in \bbZ \cup \{-\infty\}$, $b \in \bbZ \cup \{+\infty\}$ and $a \leq b$) and $\bot$ is a special element meaning ``unreachable''. $S$ and $\abstr{S}$ are ordered; here, $S$ is ordered by set inclusion $\subseteq$ and $\abstr{S}$ is ordered by $\sqsubseteq_P$, the pointwise application of $\sqsubseteq$ for all program locations: $\bot \sqsubseteq \abstr{x}$ for all $x$ in $\abstr{S}$, and
$\left((a_1, b_1), (a_2, b_2), (a_3, b_3)\right) \sqsubseteq
 \left((a'_1, b'_1), (a'_2, b'_2), (a'_3, b'_3)\right)$ if for all $1 \leq i \leq 3$, $a'_i \leq a_i$ and $b_i \leq b'_i$. For the sake of simplicity, we shall give examples further on where $P$ is a singleton; the generalization to any finite $P$ is straightforward. $P \rightarrow (\{ \bot \} \cup I^3)$ is then isomorphic to $\{ \bot \} \cup I^3$ and we shall thus consider, as a running example, the case where $S$ is $\parts{\bbZ^3}$ and $\abstr{S}$ is $\{ \bot \} \cup I^3$.

$S$ and $\abstr{S}$ are connected by an \emph{abstraction map} $\alpha$ and a \emph{concretization map} $\gamma$. $\gamma$ maps any abstract state $\abstr{x}$ to the set of concrete states that it represents. Here, $\gamma\left((a_1, b_1), (a_2, b_2), (a_3, b_3)\right)$ is the set of triples $(v_1,v_2,v_3)$ such that for all $1 \leq i \leq 3$, $a_i \leq v_i \leq b_i$. $\alpha$ maps a set $x$ of concrete states to the ``best'' (least) abstract element $\abstr{x}$ such that $x \subseteq \gamma(\abstr{x})$. Here, if $x \subseteq \bbZ^3$, then for all $1 \leq i \leq 3$, $a_i = \inf_{(v_1,v_2,v_3) \in x} v_i$ and
$b_i = \sup_{(v_1,v_2,v_3) \in x} v_i$. $\gamma$ must be monotone with respect to $\subseteq$ and $\sqsubseteq$: if $\abstr{x} \sqsubseteq \abstr{y}$, then $\gamma(\abstr{x}) \subseteq \gamma(\abstr{y})$.

In some presentations of abstract interpretation, abstract elements $\abstr{x}$ are identified with their concretization $\gamma(\abstr{x})$. For instance, one talks directly of the interval $[a,b]$, not of the pair $(a,b)$. This can make explanations smoother by clearing up notations. It is however important for some purposes to distinguish the machine representation of an abstract element $\abstr{x}$ from its concretization $\gamma(\abstr{x})$, if only because $\gamma$ may not be injective. For instance, $x=y \wedge x\leq 1$ and $x=y \wedge y \leq 1$ define exactly the same part of the plane (as geometrical convex polyhedra) but are different in their machine representation. This is the same difference as that between the \emph{syntax} and the \emph{semantics} of a logic.

In this article, we ditinguish this syntactic and semantical aspects, for several reasons. First, certain abstract operations may be sensitive to the syntax of an abstract element; that is, they may yield different results for $\abstr{x}$ and $\abstr{y}$ even though $\gamma(\abstr{x}) = \gamma(\abstr{y})$, as we shall recall in \S\ref{part:Kleene} about the polyhedra and octagons.

Also, while in many cases $\sqsubseteq$ is defined by $a \sqsubseteq b \iff \gamma(a) \subseteq \gamma(b)$, this relation may sometimes be too costly or impossible to compute, and some smaller relation may be used. For instance, if one uses a product of several abstract domains $\abstr{D}_1 \times \dots \times \abstr{D}_m$, each $D_i$ fitted with a decidable ordering $\sqsubseteq_i$, and $\gamma(\abstr{x}_1, \dots, \abstr{x}_m) = \gamma_1(\abstr{x}_1) \cap \dots \cap \gamma_m(\abstr{x}_m)$ then it is straightforward to consider the product ordering $(x_1, \dots, x_m) \sqsubseteq (x'_1, \dots, x'_m) \iff x_1 \sqsubseteq_i x'_1 \land \dots \land x_m \sqsubseteq_m x'_m$. If $x \sqsubseteq x'$ for this ordering, then $\gamma(x) \subseteq \gamma(x')$, but the two are not necessarily equivalent. Consider for instance a simplification of the domain of difference bounds \citep{mine:phd}, expressed as a product of simpler domains: the concrete states in $\bbQ^3$, the abstract domains $\abstr{D}_1 = \abstr{D}_2 = \abstr{D}_3 = \bbQ$, $\gamma_1(c_1) = \{ (x, y, z) \in \bbQ^3 \mid x - y \leq c_1 \}$, $\gamma_2(c_2) = \{ (x, y, z) \in \bbQ^3 \mid y - z \leq c_2 \}$, $\gamma_3(c_3) = \{ (x, y, z) \in \bbQ^3 \mid x - z \leq c_3 \}$. Obviously, $\gamma(1,1,2) = \gamma(1,1,3)$, yet $(1,1,3) \not\sqsubseteq (1,1,2)$. In order to use the product ordering, one has to perform beforehands a reduction operation mapping $(1,1,3$ to $(1,1,2)$, but such an operation may be nontrivial: the one in the octagon abstract domain involves a Floyd-Warshall shortest path computation, the one in the template linear constraints \citep{Sankaranarayana+others/05/Scalable} involves linear programming. In the case of real-life static analysis tools, e.g. the Astr\'ee static analyzer \cite{ASTREE_PLDI03}, with many nontrivial abstract domains interacting, it is not obvious whether $\gamma(a) \subseteq \gamma(b)$ is decidable, and even if it were, how to decide it within acceptable time.

Finally, since our goal is to write programs and proofs in a proof assistant based on intuitionistic type theory, we thought it best to clearly separate the computational, constructive content from the non-computational content: membership in the set of reachable states of a program is, in general, recursively enumerable but not recursive (from Turing's halting problem: one cannot in general decide whether the ``end'' line of the program is reachable); thus the characteristic function of that set cannot be defined by constructive logic, since this would involve describing an algorithm computing that function.

\subsection{Obtaining invariants}
Abstract interpretation replaces a possibly infinite number of concrete program execution, which cannot be simulated in practice, by a simpler ``abstract'' execution. For instance, one may replace running a program using our three integer variables over all possible initial states by a single abstract execution with interval arithmetic. The resulting final intervals are guaranteed to contain all possible outcomes of the concrete program. More formally, if one has a transition relation $\tau \subseteq \Sigma \times \Sigma$, one defines the forward concrete transfer function $f_\tau: S \rightarrow S$ as $f_\tau(x) = \{ \sigma' \mid \sigma \rightarrow_\tau \sigma' \land \sigma \in x\}$. $f_\tau(W)$ is the set of states reachable in one forward step from $W$. We say that the \emph{abstract transfer function} $\abstr{f_\tau}(\abstr{x})$ is a correct abstraction for $f_\tau$ if for all $\abstr{x}$, $f_\tau \circ \gamma(\abstr{x}) \subseteq \gamma \circ \abstr{f_\tau}(\abstr{x})$. This \emph{soundness property} means that if we have a superset of the concrete set of states before the execution of $\tau$, we get a superset of the concrete set of states after the execution of $\tau$.

As usual in program analysis, obtaining loop invariants is the hardest part. Given a set $x_0 \subseteq \Sigma$ of initial states, we would like to obtain a superset of the set of reachable states $x_\infty = \{ \sigma' \mid \sigma \rightarrow_\tau^* \sigma' \land \sigma \in x_0 \}$. The set of states $x_n$ reachable in at most $n$ steps from $x_0$ is defined by induction: $x_{n+1} = \phi(x_n)$, where $\phi(x) = f_\tau(x) \cup x_0$ is monotone, because $f_\tau$ is by definition a $\cup$-morphism. The sequence $(x_n)$ is ascending, and its limit is $x_\infty$, which is the least fixed point of $\phi$ by Kleene's fixed point theorem; this sequence is thus often known as \emph{Kleene iterations}. $x_\infty$ is also known as the \emph{strongest invariant} of the program. An \emph{inductive invariant} or \emph{post-fixpoint} is a set $x$ such that $x_0 \subseteq x$ and $f_\tau(x) \subseteq x$, and by Tarski's theorem, the intersection of all such sets is~$x_\infty$.

Obviously, the set of all possible states (often noted $\top$) is an inductive invariant, but it is uninteresting since it cannot be used to prove any non-trivial property of the program. A major goal of program analysis is to obtain program invariants $x$ that are strong enough to prove interesting properties, without being too costly to establish.

In some cases, interesting inductive invariants may be computed directly. Various approaches have recently been proposed for the direct computation of invariants, without Kleene iterations.  \citet{DBLP:conf/cav/CostanGGMP05} proposed a method for computing least fixed points in the lattice of real intervals by downward \emph{policy iteration}, also known as \emph{strategy iteration}, a technique borrowed from game theory; they later extended their framework to other domains. \citet{Gawlitza_Seidl_ESOP07} proposed a method for computing least fixed points in certain lattices by upward strategy iteration. \citet{Monniaux_SAS07,Monniaux_POPL09} showed that least fixed point problems in some lattices expressing numerical constraints can be reduced to \emph{quantifier elimination} problems, which in turn can be solved algorithmically. Other recent proposals include expressing the least invariant problem in the abstract lattice directly as a constrained minimization problem, then solving it with operational research tools~\cite{Cousot05-VMCAI}.
One common factor to these approaches is that they target specific classes of abstract domains and programs; in addition, they may also suffer from high complexity.

\subsection{Abstract Kleene iterations and widening operators}
\label{part:Kleene}
The more traditional approach to finding inductive invariants by abstract interpretation is to perform \emph{abstract Kleene iterations}.
Let $\abstr{x}_0$ be an abstraction of $x_0$. Define $\abstr{\phi}(\abstr{x}) = \abstr{f}_\tau(\abstr{x}) \sqcup \abstr{x}_0$, where $\sqcup$ is a sound overapproximation of the concrete union $\cup$: $\gamma(\abstr{x}) \cup \gamma(\abstr{y}) \subseteq \gamma(\abstr{x} \sqcup \abstr{y})$. From the soundness of $\abstr{f}_\tau$ and $\sqcup$, $\abstr{\phi}$ is a sound abstraction of $\phi$: for all $\abstr{x}$, $\phi \circ \gamma (\abstr{x}) \subseteq \gamma \circ \abstr{\phi} (\abstr{x})$. By induction, for all $n$, $x_n \subseteq \gamma(\abstr{x}_n)$: assuming $x_n \subseteq \phi(\abstr{x}_n)$, $x_{n+1} = \phi(x_n) \subseteq \phi \circ \gamma(\abstr{x}_n) \subseteq \gamma \circ \abstr{\phi}(\abstr{x}_n) = \abstr{x}_{n+1}$.

In many presentations of abstract interpretation, it is supposed that the abstract transfer function $\abstr{f}_\tau$ and the abstract union $\sqcup$ are monotonic. Intuitively, this means that if the analysis has more precise information at its disposal, then its outcome is more precise. This is true for elementary transfer functions in most abstract domains, and thus of their composition into abstract transfer functions of more complex program constructions. A well-known exception is when the abstract transfer function is itself defined as the overapproximation of a least fixed-point operation using a widening operator (see below), yet there exist less well-known cases where the abstract transfer function may be non-monotonic.%
\footnote{Such is for instance the case of the symbolic constant propagation domain proposed by \citet[\S 5]{mine:vmcai06}\citep[\S 6.3.4]{mine:phd}. The full symbolic propagation strategy can induce non-monotonic effects: if the analysis knows more relationships, it can perform spurious rewritings and paradoxically provide a less precise result.

The same is true of \citeauthor{mine:phd}'s linearization step, which dynamically abstracts nonlinear expressions as linear expressions. Consider the nonlinear expression $x \times y$ where $x \in [m_x,M_x]$, $y \in [m_y, M_y]$ and $m_x, m_y > 0$: a choice has to be made between several valid linearizations, here $x \times [m_y, M_y]$ and $[m_x,M_x] \times y$. While all choices between candidate linearizations lead to sound results, they do not have the same precision and the choice heuristic does not necessarily choose the one leading to the most precise results later on.}

Let us nevertheless temporarily assume that $\abstr{f}_\tau$ and $\sqcup$ and, thus, $\abstr{\phi}$, are monotonic, and that $\abstr{a},\abstr{b} \sqsubseteq \abstr{a} \sqcup \abstr{b}$ for all $\abstr{a}$ and~$\abstr{b}$. Then $\abstr{x_0} \sqsubseteq \abstr{x_1}$ and by induction, for all $n$, $\abstr{\phi}$ being monotonic, $\abstr{x}_n = {\abstr{\phi}}^n(\abstr{x_0}) \sqsubseteq  {\abstr{\phi}}^n(\abstr{x_1}) = \abstr{x}_{n+1}$; the sequence $\abstr{x}_n$ is therefore ascending. If this sequence is stationary, there is a $N$ such that $\abstr{x}_{N+1} = \abstr{x}_N$. Then, $\gamma(\abstr{x}_N) = \gamma(\abstr{x}_{N+1}) = \gamma(\abstr{f}_\tau(\abstr{x}_N) \sqcup \abstr{x}_0) \supseteq \gamma \circ \abstr{f}_\tau(\abstr{x}_N) \supseteq f_\tau \circ \gamma (\abstr{x}_N)$, and $\gamma(\abstr{x}_N) = \gamma(\abstr{x}_{N+1}) = \gamma(\abstr{f}_\tau(\abstr{x}_N) \sqcup \abstr{x}_0) \supseteq \gamma(\abstr{x}_0)$, which means that $\gamma(\abstr{x}_N)$ is an inductive invariant. Obviously, if the abstract domain $\abstr{S}$ is finite, then any ascending sequence is stationary.%
\footnote{This explains the popularity of Boolean abstractions: $\abstr{S}$ is the set of sets of bit vectors of fixed length $L$, and these sets are often represented by \emph{reduced ordered binary decision diagrams} (ROBDD)~\cite{ClarkeGrumbergPeled99}. Reachability analysis in BDD-based model-checkers is thus a form of Kleene iteration in the BDD space. Very astute implementation techniques, involving generalized hashing of data structures, ensure that equality tests take constant time and that $\abstr{\phi}$ is computed efficiently.}

More generally, the same results hold for any domain of \emph{finite height} (there exists an integer $L$ such that any strictly ascending sequence has at most length $L$), and, even more generally, for any domain satisfying the \emph{ascending chain condition} (there does not exist any infinite strictly ascending sequence). Yet, even the very simple domain of products of intervals that we defined earlier does not satisfy the ascending chain condition!

In domains that do not satisfy the ascending condition, the abstract Kleene iterations may fail to converge in finite time. Such is the case, for instance, of the interval abstraction of the program with a single integer variable defined by the transition system $\tau$: for all $n$, $n \rightarrow_\tau n+1$, and the initial state is~$0$. The best abstract transfer function $\abstr{\phi}$ maps a pair $(0,n)$ representing an integer interval $\{0,\dots,n\}$ to the pair $(0,n+1)$, thus the abstract Kleene iterations are $\abstr{x}_n = (0,n)$ and the analysis fails to converge in finite time.

The traditional solution to the convergence problem in domains that do not satisfy the ascending chain condition is to use a \emph{widening operator}, which is a form of convergence accelerator applied to abstract Kleene iterations \cite[Def.~4.1.2.0.4]{Cousot_state_doctorate}\cite[\S 4]{CousotCousot_JLC92}.
Intuitively, the widening operation examines the first abstract Kleene iterations and conjectures some possible over-approximation of the limit, which is then checked for stability; further iterations may be necessary until an inductive invariant is reached.
For each infinite height domain, one or more widening operators must be designed. Consequently, most literature on abstract interpretation domains includes descriptions of widening operators.

For instance, the interval abstract domain can be fitted with a simple widening discarding unstable bounds \citep{Cousot_state_doctorate}, then later with the less brutal ``widening up to'' \citep[\S 3.2]{Halbwachs_CAV93} or ``widening with thresholds''\citep[\S 6.4]{BlanchetCousotEtAl02-NJ}\cite[\S 7.1.2]{ASTREE_PLDI03}. The domain of convex polyhedra was first fitted with a very simple widening that discarded all unstable constraints \citep{CousotHalbwachs78}, but this widening was later refined in order to make it insensitive to syntactic variations in the way semantically equivalent constraints were given \citep[p.~56--57]{Halbwachs_PhD79}\citep[\S 2.2]{Halbwachs_CAV93}. \citet{mine:phd} fitted the octagon abstract domain with a similar construction, widening to $+\infty$ the unstable constraints. Again, this widening was sensitive to syntax, which lead to proposals of semantic widenings \cite{BagnaraHMZ05}. Widening techniques are not restricted to numerical domains; for instance there are specific techniques for widening over automata~\cite{DSilva_MS} (roughly speaking, they overapproximate a language defined by an automaton by the language defined by a quotient, of limited size, of that automaton; the limited size ensures termination).

Here is the most common definition:
\begin{defn}\label{defn:widening1}
A widening operator $\widening$ on an abstract domain $(\abstr{S}, \sqsubseteq)$ is a binary operator that satisfies the three following properties:
\begin{enumerate}
\item $\abstr{x} \sqsubseteq \abstr{x} \widening \abstr{y}$
\item $\abstr{y} \sqsubseteq \abstr{x} \widening \abstr{y}$
\item for any sequence $\abstr{v}_n$, a sequence of the form $\abstr{u}_{n+1} = \abstr{u}_n \widening \abstr{v}_n$ is ultimately stationary.
\end{enumerate}
\end{defn}

We can then use $\abstr{u}_0 = \abstr{x}_0$, $\abstr{u}_{n+1} = \abstr{u}_n \widening \abstr{\phi}(\abstr{u}_n)$. By the third property of the widening operator, there exists $N$ such that $\abstr{u}_N = \abstr{u}_N \widening \abstr{\phi}(\abstr{u}_N)$. Thus, $\abstr{\phi}(\abstr{u}_N) \sqsubseteq \abstr{u}_N$, and $\gamma \circ \abstr{\phi}(\abstr{u}_N) \subseteq \gamma(\abstr{u}_N)$. But $x_0 \cup f_\tau \circ \gamma (\abstr{u}_N) = \phi \circ \gamma(\abstr{u}_N) \subseteq \gamma \circ \abstr{\phi}(\abstr{u}_N) \subseteq \gamma(\abstr{u}_N)$ thus $f_\tau \circ \gamma (\abstr{u}_N) \subseteq \gamma(\abstr{u}_N)$ and $\gamma(\abstr{u}_N)$ is an inductive invariant.

Let us now have a second look at the hypotheses that we used to establish that result. Though it is often assumed that the abstract domain is a complete lattice, and that the abstract transfer function is monotonic,  we never used either hypotheses. In fact, the only hypotheses that we used are:
\begin{itemize}
\item $f_\tau$ is monotonic and the concrete domain $\parts{S}$ is a complete lattice, thus $\phi$ has a least fixed point which is the least inductive invariant of the program.
\item For all $\abstr{a}$ and $\abstr{b}$, $\abstr{b} \sqsubseteq \abstr{a} \widening \abstr{b}$.
\item For all sequence $\abstr{v}_n$, any sequence defined by $\abstr{u}_{n+1} = \abstr{u}_n \widening \abstr{v}_n$ is stationary.
\end{itemize}

\section{Relaxation of conditions and interpretation in inductive types}
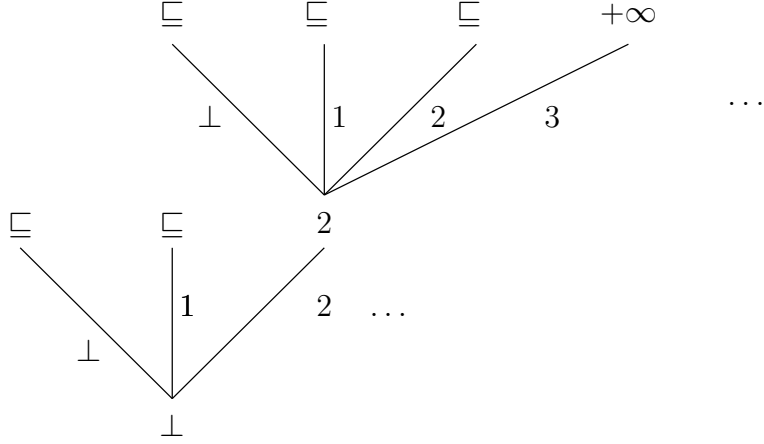
\begin{figure}
\begin{center}
\input{arbre_widening.pstex_t}
\end{center}
\caption{Interpretation of widening as a well-founded tree for the domain $1 \sqsubset 2 \sqsubset 3 \sqsubset \dots +\infty$. This domain may be used to construct the domain of intervals: an interval $[x,y]$ is represented by the pair $(-x,y) \in \bbN^2$, pointwise ordered, and the widening operation described here is applied to each coordinate.
Each node represents a proposal $\abstr{u}_n$ from the widening system. Each edge is labelled with the answer $\abstr{v}_n$ from the analysis system. The widening system either answers $\sqsubseteq$ when it determines that $\abstr{v}_n \sqsubseteq \abstr{u}_n$, or makes a new proposal. A proposal of $+\infty$ forces termination: whatever $\abstr{u}_n$ the analysis system then supplies, $\abstr{u}_n \sqsubseteq +\infty$ (we left out its outgoing branches, all finishing in $\sqsubseteq$). A path from the root of the tree is an abstract Kleene iteration sequence. The well-foundedness of the tree ensures the termination of such sequences.}
\label{fig:tree}
\end{figure}

During our work on the Astr\'ee tool \citep{ASTREE_PLDI03}, and when formalizing the notion of widening in the Coq proof assistant \citep{CoqArt},%
\footnote{Coq is a proof assistant based on higher order logic, available from \url{http://coq.inria.fr}.}
we realized that the usual definitions of abstract domains and widenings are unnecessarily restrictive for practical purposes. \citet[\S4.4]{Pichardie_PhD} already proposed a relaxation of these conditions, but his definition of widenings is still fairly complex. We propose here a drastically reduced informal definition of widenings, which subsumes both the $\sqsubseteq$ ordering and the $\widening$ operator; this definition will be made formal as Def.~\ref{defn:new_widening2}.

\begin{defn}\label{defn:new_widening}
A widening system is an algorithm that proposes successive abstract elements $\abstr{u}_0, \abstr{u}_1, \dots, \abstr{u}_n$ to the rest of the analyzer, and receives $\abstr{v}_n$ from it. It can then either terminate with some guarantee that $\gamma(\abstr{v}_n) \subseteq \gamma(\abstr{u}_n)$, or propose the next element $\abstr{u}_{n+1}$. The system never provides infinite sequences.
\end{defn}

In practical use, $\abstr{v}_n = \abstr{\phi}(\abstr{u}_n)$ and $\abstr{\phi}$ is an abstraction of the concrete transformer $\phi$ of a loop or, more generally, of a monotonic system of semantic equations.

It is obvious that any widening that verifies the conditions of Def.~\ref{defn:widening1} also verifies these conditions. Note that Def.~\ref{defn:new_widening} is strictly laxer than Def.~\ref{defn:widening1}. For instance, we make no requirement that $\gamma(\abstr{u}_n) \subseteq \gamma(\abstr{u}_{n+1})$; a widening system could first try some ascending sequence $\abstr{u}_0, \dots, \abstr{u}_n$, regret, and restart with another sequence $\abstr{u}_{n+1}, \dots$.

A more mathematical way of seeing this definition is by interpreting the widening system as a well-founded tree:

\begin{defn}\label{defn:new_widening2}
Let $\abstr{S}$ be an abstract domain with the associated concretization map $\gamma$. Let $\sqsubseteq$ be a preorder over $\abstr{S}$ such that $\gamma$ is monotonic. A widening system is a well-founded tree whose nodes are labeled by elements of $\abstr{S}$ (there may be several nodes with the same label). From a node labeled with $\abstr{u}$, there are branches labeled with every $\abstr{v}$ such that $\abstr{v} \not \sqsubseteq \abstr{u}$.
\end{defn}

Let $\abstr{u}_0$ be the label for the root of the tree, and let $\abstr{u}_0, \abstr{v}_0, \abstr{u}_2, \dots$ be a path into the tree consisting in successive nodes and edges. Because the tree is well-founded, this path is finite, which means that it terminates with $\abstr{u}_N, \abstr{v}_N$ such that $\abstr{v}_N \sqsubseteq \abstr{u}_N$. This recalls the termination property of Def.~\ref{defn:widening1}.

Definition~\ref{defn:new_widening2}, combined with the $\sqsubseteq$ test can be easily recast as couple of mutually inductive types~:

\begin{equation}
\begin{array}{ll}
\textit{widening} & \equiv \abstr{S} \times (\abstr{S} \rightarrow \textit{answer})\\
\textit{answer} & \equiv \textit{termination} \mid \textit{next} \textrm{~of~} \textit{widening}\\
\end{array}
\end{equation}

From each node labeled by $\abstr{u}$, for each $\abstr{v}$ there is an edge labeled by $\abstr{v}$, which either leads to ``termination'' if $\abstr{v} \sqsubseteq \abstr{u}$, or to another node (see Fig.~\ref{fig:tree}).

Note that, even in an eager language such as Objective Caml, the widening tree is never constructed in memory: its nodes are constructed on demand by application of the function $\abstr{S} \rightarrow \textit{answer}$.

In a higher-order type system with dependent sums and products such as the Calculus of inductive constructions (as in Coq), the above inductive datatype can be adorned with proof terms. A tree node $\textit{widening}$ is a pair $(\abstr{u}, a)$ where $a$ maps each $\abstr{v}$ to an \emph{answer}. $a(\abstr{v})$ is either ``$\sqsubseteq$'', carrying a proof term stating that $\gamma(\abstr{v}) \subseteq \gamma(\abstr{u})$, or another $\textit{widening}$ tree node.

\section{Implementation in Coq}
We shall first show how to implement our concept of widening system in Coq, then we shall give a few concrete examples of how common abstract interpretation techniques can be implemented within this framework.%
\footnote{Source code may be downloaded from\\
\url{http://www-verimag.imag.fr/~monniaux/download/domains_coq.zip}.}

\subsection{Framework}
 We assume we have an abstract domain \texttt{S} with an ordering \texttt{domain\_le} (representing $\sqsubseteq$). In practice, this ordering is supposed to be decidable: there exists a function \texttt{domain\_le\_decide} that takes $x$ and $y$ as inputs and decides whether $x \sqsubseteq y$.

The \emph{answer} is the disjunctive sum \verb@{domain_le y x} + widening@: it either provides a new \texttt{widening} object, or a proof that $y \sqsubseteq x$. By inlining this type into the definition of \texttt{widening}, we obtain:

\begin{verbatim}
Variable S : Set.
Hypothesis domain_le : S -> S -> Prop.
Hypothesis domain_le_decide :
    forall x y : S,
      { domain_le x y } + {~ (domain_le x y) }.

Inductive widening: Set :=
  widening_intro : forall x : S,
    (forall y : S, widening + {domain_le y x}) -> widening.
\end{verbatim}

Note that all properties desired of the widening are lumped in this definition. The \texttt{Inductive} keyword introduces a type whose elements are all well-founded by construction; Coq will make it impossible to create widening trees that are not well-founded. The correct termination property (termination only if $\abstr{v} \sqsubseteq \abstr{u}$) is also ensured by construction: a leaf edge corresponding to $\abstr{u}$ and $\abstr{v}$ may be constructed only by giving a proof of $\abstr{v} \sqsubseteq \abstr{u}$ (a term belonging to the type \texttt{domain\_le} $\abstr{v}$ $\abstr{u}$).

In the above definition, we have added the hypothesis that $\sqsubseteq$ is decidable (\texttt{domain\_le\_decide}). This is not needed for this definition, but is useful in many constructions, and is a very reasonable assumption to make. Indeed, the reason why we introduced $\sqsubseteq$ as just any order such that $\gamma$ is monotonic, and not the most precise one, is that the most precise one might not be decidable, or too costly to decide effectively.

Since \texttt{widening} is an inductive type, defining well-founded trees, it is possible to define functions by induction over elements of that type. One especially interesting inductively defined function takes $\abstr{f}: \abstr{S} \rightarrow \abstr{S}$ as a parameter and computes $\abstr{x}$ such that $\abstr{f}(\abstr{x}) \sqsubseteq \abstr{x}$ by well-founded induction over the widening tree.
On a widening node labeled by $\abstr{u}$, it computes $\abstr{v}=\abstr{f}(\abstr{u})$ then requests the ``answer'' from the widening node on the value $\abstr{v}$:
\begin{itemize}
\item Either it answers with another widening node and the function is called recursively.
\item Or it answers with a proof that $\abstr{v} \sqsubseteq \abstr{u}$ and the algorithm terminates with the requested answer (both $\abstr{u}$ and a proof that $\abstr{f}(\abstr{u}) \sqsubseteq \abstr{u})$.
\end{itemize}

\begin{verbatim}
Section Recursor.
Variable f : S -> S.

Fixpoint abstract_lfp_rec
  (iteration_step : widening) :
  { lfp : S | domain_le (f lfp) lfp } :=
  let (x, xNext) := iteration_step in
  match xNext (f x) with
  | inleft next_widening => abstract_lfp_rec next_widening
  | inright fx_less_than_x => exist (fun x => domain_le (f x) x)
      x fx_less_than_x
  end.

End Recursor.
\end{verbatim}

For ease of use, we pack \texttt{S}, \texttt{domain\_le}, an abstraction relation \texttt{domain\_abstracts} and other related constructs into one single \texttt{domain} record.

\subsection{Examples}
In numerical abstract domains, it is common to use ``widening up to'' \cite[\S 3.2]{Halbwachs_CAV93} or ``widening with thresholds''~\citep[\S 6.4]{BlanchetCousotEtAl02-NJ}\cite[\S 7.1.2]{ASTREE_PLDI03}: one keeps an ascending sequence $\abstr{z}_1, \dots, \abstr{z}_n$ of ``magical'' values, and $\abstr{x} \widening \abstr{y}$ is the least element $\abstr{z}_k$ greater than $\abstr{x} \sqcup \abstr{y}$. For instance, instead of widening a sequence of integer intervals $[0,1]$, $[0,2]$ etc. to $[0,+\infty[$, we may try some ``magical'' values such as $[0,255]$, $[0,32767]$ etc. Yet, if all elements in the sequence fail to define an inductive invariant, we are forced to try $[0,+\infty[$. In other words, after trying the ``magical'' values, we revert to the usual brutal widening on the intervals.

This is easily modeled within our framework by a ``widening transformer'': taking a widening $W$ as input and a finite ``ramp'' $l$ of values, it outputs a widening $W'$ that first applies the thresholds and, as a last resort, calls~$W$. \texttt{Variable T : domain} is a parameter defining the original domain and original widening, which is used as the last resort by our transformed widening. Function \texttt{ramp\_widening\_search} searches for the next threshold in the~``ramp''.

\begin{verbatim}
Section Widening_ramp.
Variable T : domain.

Fixpoint ramp_widening_search (bound : (domain_set T))
  (ramp : (list (domain_set T))) { struct ramp } : (list (domain_set T)) :=
  match ramp with
  | nil => ramp
  | (cons ramp_h ramp_t) =>
    match (domain_le_decide T bound ramp_h) with
    | left _ => ramp
    | right _ => ramp_widening_search bound ramp_t
    end
  end.

Fixpoint ramp_widening (ramp : (list (domain_set T))) :
  (widening (domain_set T) (domain_le T)) :=
  match ramp with
  | nil => domain_widening T
  | (cons ramp_h ramp_t) =>
    (widening_intro (domain_set T) (domain_le T) ramp_h
      (fun (y : (domain_set T)) =>
       match domain_le_decide T y ramp_h with
       | left STOP =>
           inright
             (widening (domain_set T) (domain_le T)) STOP
       | right _ =>
          inleft
             (domain_le T y ramp_h)
             (ramp_widening (ramp_widening_search y ramp_t))
       end))
  end.
\end{verbatim}

A trick often used in static analysis is to delay the widening~\cite[\S 7.1.3]{ASTREE_PLDI03}. Instead of performing $\widening$ at each iteration, one performs $\sqcup$ for a finite number of steps, then one tries $\widening$ again. For termination purposes, it suffices that there is some ``fairness property'': $\widening$ should not be delayed infinitely. 
One can for instance choose to delay widening by $n$ steps of $\sqcup$ after each widening step. This is again implemented as a ``widening transformer'':
\begin{verbatim}
Definition delayed_widening_each_step :
  nat -> (widening (domain_set T) (domain_le T)).
\end{verbatim}

We can similarly build a product domain $\abstr{S}_1 \times \abstr{S}_2$. The widening on couples $(a_1,a_2) \widening (b_1,b_2) = (a_1 \widening_1 b_1, a_2 \widening_2 b_2)$ is implemented by a ``widening transformer'' taking one widening $W_1$ on $\abstr{S}_1$ and a widening $W_2$ on $\abstr{S}_2$ as inputs, and producing a widening on $\abstr{S}_1 \times \abstr{S}_2$ by syntactic induction on $W_1$ and $W_2$: if $a_1 \sqsubseteq_1 b_1 \land a_2 \sqsubseteq_2 b_2$, then $(a_1,a_2) \sqsubseteq (b_1,b_2)$ for the product ordering and one terminates; if $a_1 \sqsubseteq_1 b_1$ but $a_2 \not\sqsubseteq_2 b_2$ then one stays on $a_1$ but moves one step into $W_2$ (and \emph{mutatis mutandis} reversing the coordinates); if $a_1 \not\sqsubseteq_1 b_1$ and $a_2 \not\sqsubseteq_2 b_2$, then one moves into both $W_1$ and $W_2$. This implements the usual widening on products. This construct can be generalized to any finite products of domains.

\section{Conclusion}
By seeing the combination of the computational ordering $\sqsubseteq$ and the widening operator $\widening$ as a single inductive construct, one obtains an elegant characterization extending the usual notion of widening in abstract interpretation, suitable for implementation in higher order logic.

\section*{Acknowledgments}
The author would like to thank the anonymous referees, whose suggestions
greatly improved this article. This work was partially funded by ANR
project ``\href{http://asopt.inrialpes.fr/index.php/Main_Page}{ASOPT}''.

\newcommand{\doix}[1]{\href{http://dx.doi.org/#1}{#1}\endgroup}
\newcommand{\doi}{\begingroup\footnotesize doi: \catcode`\_=13\def\_{\textunderscore}\doix}
\bibliographystyle{plainnat}
\bibliography{minimalistic_widening}
\end{document}

%% file: arbre_widening.pstex_t
\begin{picture}(0,0)%
\includegraphics{arbre_widening.pstex}%
\end{picture}%
\setlength{\unitlength}{4144sp}%
\begingroup\makeatletter\ifx\SetFigFontNFSS\undefined%
\gdef\SetFigFontNFSS#1#2#3#4#5{%
  \reset@font\fontsize{#1}{#2pt}%
  \fontfamily{#3}\fontseries{#4}\fontshape{#5}%
  \selectfont}%
\fi\endgroup%
\begin{picture}(4215,2710)(2236,-7100)
\put(2656,-6586){\makebox(0,0)[b]{\smash{{\SetFigFontNFSS{12}{14.4}{\rmdefault}{\mddefault}{\updefault}{\color[rgb]{0,0,0}$\bot$}%
}}}}
\put(3241,-6316){\makebox(0,0)[b]{\smash{{\SetFigFontNFSS{12}{14.4}{\rmdefault}{\mddefault}{\updefault}{\color[rgb]{0,0,0}1}%
}}}}
\put(4051,-6316){\makebox(0,0)[b]{\smash{{\SetFigFontNFSS{12}{14.4}{\rmdefault}{\mddefault}{\updefault}{\color[rgb]{0,0,0}2}%
}}}}
\put(4321,-6316){\makebox(0,0)[lb]{\smash{{\SetFigFontNFSS{12}{14.4}{\rmdefault}{\mddefault}{\updefault}{\color[rgb]{0,0,0}$\dots$}%
}}}}
\put(4051,-5821){\makebox(0,0)[b]{\smash{{\SetFigFontNFSS{12}{14.4}{\rmdefault}{\mddefault}{\updefault}{\color[rgb]{0,0,0}2}%
}}}}
\put(3151,-7036){\makebox(0,0)[b]{\smash{{\SetFigFontNFSS{12}{14.4}{\rmdefault}{\mddefault}{\updefault}{\color[rgb]{0,0,0}$\bot$}%
}}}}
\put(3376,-5191){\makebox(0,0)[b]{\smash{{\SetFigFontNFSS{12}{14.4}{\rmdefault}{\mddefault}{\updefault}{\color[rgb]{0,0,0}$\bot$}%
}}}}
\put(3241,-6316){\makebox(0,0)[b]{\smash{{\SetFigFontNFSS{12}{14.4}{\rmdefault}{\mddefault}{\updefault}{\color[rgb]{0,0,0}1}%
}}}}
\put(4141,-5191){\makebox(0,0)[b]{\smash{{\SetFigFontNFSS{12}{14.4}{\rmdefault}{\mddefault}{\updefault}{\color[rgb]{0,0,0}1}%
}}}}
\put(4726,-5191){\makebox(0,0)[b]{\smash{{\SetFigFontNFSS{12}{14.4}{\rmdefault}{\mddefault}{\updefault}{\color[rgb]{0,0,0}2}%
}}}}
\put(2251,-5821){\makebox(0,0)[b]{\smash{{\SetFigFontNFSS{12}{14.4}{\rmdefault}{\mddefault}{\updefault}{\color[rgb]{0,0,0}$\sqsubseteq$}%
}}}}
\put(3151,-4561){\makebox(0,0)[b]{\smash{{\SetFigFontNFSS{12}{14.4}{\rmdefault}{\mddefault}{\updefault}{\color[rgb]{0,0,0}$\sqsubseteq$}%
}}}}
\put(4006,-4561){\makebox(0,0)[b]{\smash{{\SetFigFontNFSS{12}{14.4}{\rmdefault}{\mddefault}{\updefault}{\color[rgb]{0,0,0}$\sqsubseteq$}%
}}}}
\put(4906,-4561){\makebox(0,0)[b]{\smash{{\SetFigFontNFSS{12}{14.4}{\rmdefault}{\mddefault}{\updefault}{\color[rgb]{0,0,0}$\sqsubseteq$}%
}}}}
\put(6436,-5056){\makebox(0,0)[lb]{\smash{{\SetFigFontNFSS{12}{14.4}{\rmdefault}{\mddefault}{\updefault}{\color[rgb]{0,0,0}$\dots$}%
}}}}
\put(5401,-5191){\makebox(0,0)[b]{\smash{{\SetFigFontNFSS{12}{14.4}{\rmdefault}{\mddefault}{\updefault}{\color[rgb]{0,0,0}3}%
}}}}
\put(5851,-4561){\makebox(0,0)[b]{\smash{{\SetFigFontNFSS{12}{14.4}{\rmdefault}{\mddefault}{\updefault}{\color[rgb]{0,0,0}$+\infty$}%
}}}}
\put(3151,-5821){\makebox(0,0)[b]{\smash{{\SetFigFontNFSS{12}{14.4}{\rmdefault}{\mddefault}{\updefault}{\color[rgb]{0,0,0}$\sqsubseteq$}%
}}}}
\end{picture}%